\documentclass[onecolumn,showpacs]{revtex4}

\usepackage{graphicx}

\newcommand\be{\begin{equation}}
\newcommand\ee{\end{equation}}
\newcommand\bea{\begin{eqnarray}}
\newcommand\eea{\end{eqnarray}}
\newcommand\nn{\nonumber}

\newcommand\psl{p \kern-.45em{/}}
\newcommand\qsl{q \kern-.5em{/}}

\begin{document}


\title{Baryonic contributions to the dilepton spectrum 
       of nucleon-nucleon collisions}

\author{M.~Z\'et\'enyi}\email{zetenyi@rmki.kfki.hu} 
\author{Gy.~Wolf}\email{wolf@rmki.kfki.hu}
\affiliation{KFKI Research Institute for Nuclear and
Particle Physics, H-1525 Budapest 114, POB. 49, Hungary}

\date{\today}

\begin{abstract}
We study the production of dileptons in relativistic nucleon-nucleon
collisions. Additionally to the traditional dilepton production
channels (vector meson decays, meson and $\Delta(1232)$ Dalitz decays)
we included in our model as new dilepton sources the Dalitz decay of
higher unflavored baryon resonances with spin$\le$5/2 and
mass$\le$2.25~GeV/c$^2$. The contributions of these 
new channels are estimated using experimental information about the
$N\gamma$ decays of the resonances and have large uncertainties.
The obtained dilepton spectra
are compared to the experimental data by the DLS collaboration.
Predictions for the HADES detector (SIS, GSI) are also discussed.
In spite of the large uncertainties of the higher resonance Dalitz decay
contributions we are able to draw the conclusion that these
contributions are negligible compared to the other dilepton sources and
do not influence the detectability of the $\phi$ and $\omega$ vector
meson peaks.
\end{abstract}

\pacs{13.75.Cs; 13.30.Ce; 13.40.Hq; 14.20.Gk}

\maketitle


\section{Introduction}

The main objective of the physics program at SIS (GSI, Darmstadt)
is the study of hot and dense hadronic matter formed in
nucleus-nucleus collisions. The clearest probes for this
investigation are dileptons, which leave the interaction volume
unaffected by strong final state interactions. The purpose
of the HADES detector is to study the dilepton spectrum
produced in nucleus-nucleus, nucleon-nucleus, nucleon-nucleon, 
and pion-nucleus collisions.

Theoretical investigations predict a partial restoration
of chiral symmetry in hot and dense nuclear matter, which results
in a modification of particle properties. For an experimental study
of these phenomena the light vector mesons $\rho$, $\omega$,
and $\phi$ seem to be a suitable probe. The $\rho$ --
owing to its short lifetime -- decays inside the hot and dense
hadronic matter. Predicted in-medium modifications of $\omega$ and
$\phi$ may result in a considerable growth of their decay widths
and there is hope that they also decay inside the hot and dense
phase. On the other hand, $\rho$, $\omega$, and $\phi$ decay directly
to dileptons, which allows a clear experimental study.

Dileptonic decays of vector mesons contribute to the dilepton
invariant mass spectrum as peaks or -- in the case of the $\rho$
-- as a wider distribution. In-medium modification of vector meson
masses and decay widths may be identified in the dilepton
spectrum as a shift and widening of the corresponding peaks,
respectively.

To identify the vector mesons in the dilepton spectrum a careful
analysis of the backgrounds is needed. One of the possible sources
of background is the Dalitz-decay of baryon resonances, $N^* \to Ne^+e^-$.
These processes may be interesting also for themselves since baryon
resonance properties may also be subject to in-medium modifications.
In earlier studies only the $\Delta(1232)$ and $N(1440)$ Dalitz-decays 
have been considered in model calculations (the latter giving a
negligible contribution). On the other hand, some of the higher resonances
have a relatively high photonic decay width and are, therefore,
expected to have a non-negligible contribution to the dilepton
spectrum through their Dalitz-decay. 
The kinematical upper limit for the mass of dileptons resulting from
the Dalitz-decay of resonances rises with increasing resonance mass,
therefore higher mass resonances may contribute to the dilepton spectrum
in a higher dilepton mass range where the contribution of the $\Delta(1232)$
is already negligible. It is particularly interesting to see if some
of the higher mass resonances contribute essentially to the dilepton
invariant mass spectrum in the $\phi$ mass region since this may
influence the possible identification of the $\phi$ meson in the 
dilepton spectrum.
Some of the higher mass resonances have also higher spins. 
In studying their electromagnetic transitions special care should
be taken of their higher spin nature.

Dilepton production in proton-proton collisions with beam energies
between 1 and 5 GeV has been studied experimentally by the 
DLS collaboration \cite{DLS}, supplying an opportunity to test
models of dilepton production in elementary processes.
These models may in turn be used as an input in transport models of 
heavy ion collisions.

In this paper we discuss the Dalitz-decay of baryon resonances
up to spin-5/2. We inserted our results into a Monte Carlo model 
of relativistic proton-proton collisions and obtained the dilepton
invariant mass spectra.

The lack of experimental information on the Dalitz-decay of baryon
resonances forces one to use information on the -- often also poorly
known -- radiative decays, $N^* \to N\gamma$. This brings in several
assumptions and leads to uncertainties of the results. The resulting
uncertainty of the total dilepton invariant mass spectrum is moderate
and -- as we will see -- allows one to draw the conclusion that the
detectability of the $\phi$ and $\omega$ meson peaks is not influenced
by the baryon resonance Dalitz-decays.

In an earlier paper \cite{ourpaper} we discussed the calculation of the
Dalitz-decay of 
baryon resonances and made a comparison of the contributions of the
various possible decay amplitudes. In Sec.\ \ref{dal.dec.} of the
present paper we briefly review those results. In Sec.\ \ref{MC}
we give a short description of our Monte Carlo model for proton-proton
collisions, in Sec.\ \ref{channels} we discuss the dilepton production 
channels included in the model. In Secs.\ \ref{DLScomp}
and \ref{HADESpred} we compare our dilepton spectra with the DLS data and
give predictions for the HADES detector.
Throughout the paper we set $c=\hbar=1$.


\section{Dalitz-decay of a spin-$J$ resonance}
\label{dal.dec.}

\subsection{Kinematics and matrix elements}
Let $p^*$, $m^*$, and $\lambda^*$ denote the four-momentum, mass,
and helicity of the decaying resonance, respectively, and $p$, $m$, and
$\lambda$ the corresponding quantities for the nucleon. Let us
introduce the notation $q = p^* - p$ for the photon four-momentum, 
$M^2 = q^2$ for the virtual photon mass (= dilepton
invariant mass) squared, and $P = (p^* + p)/2$.

The differential width of the Dalitz-decay of a
particle is related to its photonic decay width to a
virtual photon $\Gamma_{N^* \to N\gamma}(M)$ by \cite{LS}
\be
\frac{d\Gamma_{N^* \to N e^+ e^-}}{dM^2} =
\frac{\alpha}{3\pi}\frac{1}{M^2}\Gamma_{N^* \to N\gamma}(M).
\ee
Here $\Gamma_{N^* \to N\gamma}(M)$ can be expressed in terms of the
photonic decay matrix element $\langle N \gamma | T | N^* \rangle$ as
\be
\Gamma_{N^* \to N\gamma}(M) =
\frac{\sqrt{\lambda ({m^*}^2,m^2,M^2)}}{16 \pi {m^*}^3}
\frac{1}{n_{pol,N^*}}
\sum_{pol} {\vert \langle N \gamma | T | N^* \rangle \vert}^2,
\label{width}
\ee
where
\be
\langle N \gamma | T | N^* \rangle =
- \epsilon^{\mu} \langle N | J_{\mu} | N^* \rangle,
\ee
with $\epsilon^{\mu}$ the photon polarization vector,
and $J_{\mu}$ the electromagnetic current operator.
In (\ref{width}) $n_{pol,N^*}$ is the number of polarization states
of the $N^*$ resonance and
$\lambda(a,b,c) = a^2 + b^2 + c^2 - 2(ab+bc+ac)$ is the usual
kinematical factor.

Next we determine the matrix element
$\langle N | J_{\mu} | N^* \rangle$ for an $N^*$ resonance of spin-$J$.
A spin-$J$ ($J \ge 3/2$) fermion can be described by a spinor-tensor field
$\Psi^{\rho_1 \cdots \rho_n}$ ($n = J - 1/2$), therefore in
momentum-space matrix elements the particle is represented by
a spinor-tensor amplitude $u^{\rho_1 \cdots \rho_n}(p^*,\lambda^*)$,
which has to fulfill the generalized Rarita-Schwinger relations in order
to eliminate the different spin degrees of freedom.
The electromagnetic current matrix element can be written generally as
\be
\langle N | J_{\mu} | N^* \rangle =
\bar{u}(p,\lambda) \Gamma_{\mu \rho_1 \cdots \rho_n}
u^{\rho_1 \cdots \rho_n}(p^*,\lambda^*),
\label{mxelement}
\ee
where the form of $\Gamma_{\mu \rho_1 \cdots \rho_n}$ is restricted
by the conservation of electric charge,
$q^{\mu} \langle N | J_{\mu} | N^* \rangle = 0$.
As a result of the Dirac equations 
$\bar{u}(p,\lambda)(\psl - m) = 0$
and $(\psl^* - m^*) u^{\rho_1 \cdots \rho_n}(p^*,\lambda^*)
= 0$ and the generalized Rarita-Schwinger relations 
$\Gamma_{\mu \rho_1 \cdots \rho_n}$ contains only three independent terms:
\be
\Gamma_{\mu \rho_1 \cdots \rho_n} =
\sum_{i=1}^{3} f_i(M^2) \chi^i_{\mu \rho_1}
p_{\rho_2} \cdots p_{\rho_n} G,
\label{gamma}
\ee
with
\bea
\chi^1_{\mu \rho} & = &
\gamma_{\mu} q_{\rho} - \qsl g_{\mu \rho}, \nn \\
\chi^2_{\mu \rho} & = &
P_{\mu} q_{\rho} - (P \cdot q) g_{\mu \rho}, \nn \\
\chi^3_{\mu \rho} & = &
q_{\mu} q_{\rho} - M^2 g_{\mu \rho},
\label{terms}
\eea
and $G=1$ or $\gamma_5$ for resonances with positive or negative
normalities, respectively. The normality of a spin-$J$ baryon is
by definition $P (-1)^{J-1/2}$ with $P$ the intrinsic parity.
$f_i(M^2)$, $i = 1,2,3$ are three independent form factors,
which should be determined from experimental data.

\subsection{Parameter fitting}
In our model we neglect the $M^2$ dependence of the form factors,
therefore $f_i$ become constant couplings.
(Vector meson dominance does not apply here since we include the
$\rho$ dileptonic decay separately.) 
We redefine the $\chi^i_{\mu\rho}$ in our matrix elements dividing them
by the appropriate power of the nucleon mass $m$ in order to
render the form factors $f_i$ dimensionless. We, further, write
$f_i = e g_i$, with $e$ the elementary charge related to the fine structure
constant by $e^2 = 4\pi\alpha$. We will use the real photonic decay
width of the resonances to determine the dimensionless coupling
constants $g_i$. 

A detailed investigation of the Dalitz-decay width calculated from the 3
independent matrix elements and the interference terms leads to the
following conclusions (see \cite{ourpaper} for details):

\begin{itemize}
\item Those terms of the Dalitz-decay width that contain
$\chi^3_{\mu\rho}$ give zero contribution at $M=0$ (real photons),
therefore information on the real photonic width of the resonances can
not be used to fix the value of the $g_3$ coupling.
\item The remaining terms ($\propto g_1^2$, $g_2^2$ and $g_1g_2$) give
very similar contributions to the Dalitz-decay width. Thus an arbitrary
choice between the $\chi^1_{\mu\rho}$ and $\chi^2_{\mu\rho}$ terms
causes only a minor uncertainty in the dilepton invariant mass spectrum,
that is definitely smaller than the uncertainty arising from the poor
knowledge of the photonic width in the case of most of the resonances.
\item The resonance mass dependence of the Dalitz-decay width may
influence the resulting dilepton spectrum because the masses of the
resonances are generated according to a Breit-Wigner distribution in our
model (see Sec.\ \ref{MC}), but we use a single coupling fitted to
the photonic width of a resonance with its mass equal to the
peak value. However, the Breit-Wigner distribution ensures that
resonance masses stay close to the peak value and hence the uncertainty
of the dilepton spectrum caused by the different resonance mass
dependence of the $\chi^1_{\mu\rho}$ and $\chi^2_{\mu\rho}$ is small.
\end{itemize}

These results encourage us to make the assumption that $g_2=g_3=0$,
i.e., only $\chi^1_{\mu\rho}$ (the term containing the lowest power of
external momenta) contributes to the Dalitz-decay of baryon resonances.

\begin{figure}[htb]
\vspace*{-0.5cm}
\begin{center}
\includegraphics[width=7.5cm]{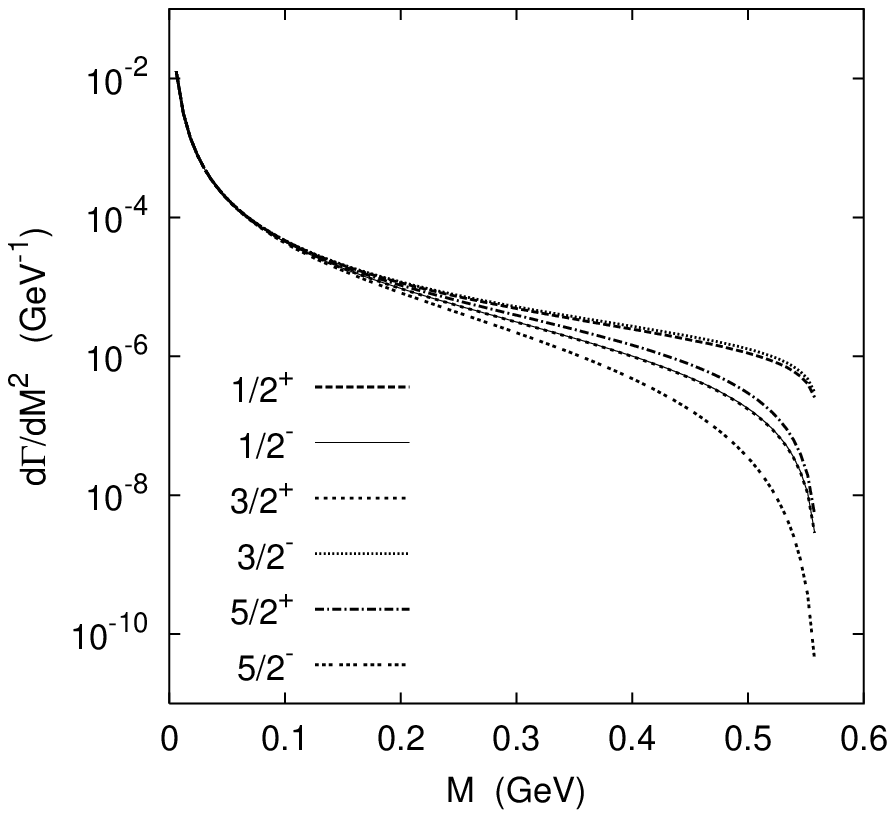}
\end{center}
\caption{\small Differential width of the Dalitz-decay of hypothetical baryon 
resonances with the same mass (1.5 GeV) and photonic width (0.6 MeV), 
but with different spin-parities. Curves for 1/2$-$ and 3/2+ resonances
nearly overlap, similarly 1/2+ and 3/2$-$ curves.}
\label{spins}
\end{figure}

To demonstrate the relevance of the spin-parity of baryon resonances
in their Dalitz-decays we show in Fig.\ \ref{spins} the differential
Dalitz-decay width of hypothetical resonances with the same mass and
photonic width but with different spin-parities. It can be seen from the 
plot that at the high dilepton mass end of the spectrum spin-parity
1/2+ and 3/2$-$ resonances give the largest while 5/2$-$ give the smallest
contribution to the dilepton spectrum.


\section{Dilepton production in proton-proton collisions}

\subsection{The model}
\label{MC}

We use a resonance model to describe particle production in proton-proton
collisions: baryon resonances are produced in the first step, which
then decay -- possibly in a multi-step process -- into the final
state particles. The resonances are treated as on-shell particles 
in their production processes, but their mass is generated according 
to a Breit-Wigner distribution.
We included in our model all unflavored baryon resonances with
mass$\le$2.25~GeV, and with a status of at least 3 stars according
to the Review of Particle Physics \cite{PDG}, i.e.,
$\Delta(1232)$, $N(1440)$, $N(1520)$, $N(1535)$,
$\Delta(1600)$, $\Delta(1620)$, $N(1650)$, $N(1675)$, $N(1680)$,
$N(1700)$, $\Delta(1700)$, $N(1710)$, $N(1720)$, $\Delta(1905)$,
$\Delta(1910)$, $\Delta(1920)$, $\Delta(1930)$, $\Delta(1950)$,
$N(2190)$, $N(2220)$, $N(2250)$,
and some additional resonances with weaker status, namely
$\Delta(1900)$, $N(2000)$, and $N(2080)$.
In our model the following decay channels of baryon resonances 
are taken into account:
$N\pi$, $N\eta$, $N\pi\pi$ (including $N\rho$, $\Delta(1232)\pi$,
$N(1440)\pi$, and $N$``$\sigma$''), $N\omega$, $\Sigma K$, and $\Lambda K$.

We fitted the resonance properties (branching ratios, full widths, 
and masses within
the errors given in the Review of Particle Physics~\cite{PDG})
to experimental data of pion induced reactions. 
Cross sections of resonance production have been fitted to
one-pion, two-pion, $\eta$, $\omega$, $\rho$, and kaon production
in proton-proton and proton-neutron collisions. 
Formulae for resonance and meson production cross sections used in the 
fits are essentially the same as in the BUU model established by Gy.\ Wolf
(see \cite{BUU} and references therein).
We needed to include in the model the resonances with weaker status 
mentioned above in order to obtain a good fit.
For further details of parameter fitting see~\cite{fit}.

\subsection{Dilepton production}
\label{channels}

We used the above model to calculate the dilepton invariant mass
spectrum of proton-proton collisions. 
The dilepton sources considered are the
direct dilepton decay of $\rho^0$, $\omega$, and $\phi$ vector mesons
and Dalitz-decay of $\pi^0$, $\eta$, and $\omega$ mesons,
and baryon resonances.
For the meson Dalitz-decay channels we used the same expressions as
in \cite{BUU}.

The differential cross section of dilepton production in the direct decay
of a $V$ vector meson is given by
\be
{\frac{d\sigma(s,M)}{dM}}^{pp\to ppV\to ppe^+e^-} =
{\frac{d\sigma(s,M)}{dM}}^{pp\to ppV} BR_{V\to e^+e^-}(M).
\label{Vdilep}
\ee
Here $BR_{V\to e^+e^-}(M)$ is the dileptonic branching ratio of the
vector meson $V$, which is approximated in the case of the $\rho$ as
\be
BR_{\rho\to e^+e^-}(M)=
\frac{\Gamma_{\rho\to e^+e^-}(M)}
{\Gamma_{\rho\to\pi\pi}(M) + \Gamma_{\rho\to e^+e^-}(M) +
\Gamma_{\rho\to\pi\gamma}(M)},
\label{BR}
\ee
where we adopt the mass dependence of the pionic and dileptonic $\rho$
width from \cite{giessen1} while we keep $\Gamma_{\rho\to\pi\gamma}$
constant, determining its value from the Breit-Wigner with and 
branching ratio listed in the Review of Particle Physics \cite{PDG}.
The appearance of $\Gamma_{\rho\to e^+e^-}$ and $\Gamma_{\rho\to\pi\gamma}$
in the denominator of (\ref{BR}) is only relevant close to the two pion 
threshold $M=2m_\pi$, where $\Gamma_{\rho\to\pi\pi}$ becomes small.

We keep the dileptonic branching ratio of the narrow $\omega$ and 
$\phi$ resonances constant and use their PDG values. $\phi$ meson
production is not handled by our resonance model and the production
cross section is not well known experimentally. This leads us
to adopt here the $\phi$ meson production cross section calculated from
a one-boson exchange model \cite{Chung}.

We calculate the dilepton spectra from the Dalitz-decay of those of
the above mentioned baryon resonances that have spin$\le$5/2 and are
known to couple to the $N\gamma$ channel according to the PDG, i.e.,
$\Delta(1232)$, $N(1440)$, $N(1520)$, $N(1535)$,
$\Delta(1600)$, $\Delta(1620)$, $N(1650)$, $N(1675)$, $N(1680)$,
$N(1700)$, $\Delta(1700)$, $N(1710)$, $N(1720)$, $\Delta(1905)$,
$\Delta(1910)$, $\Delta(1930)$.
For resonances with spin$\ge$3/2 we use the results of the previous
section. The Dalitz-decay of the $N(1440)$ resonance has already been
included in Ref.\ \cite{N1440}. We treated all the other spin-1/2
resonances in the same fashion.

To fix the electromagnetic coupling constants $g_1$ defined in 
Sec.\ \ref{dal.dec.} the photonic width of the resonances is
needed. This we calculate using the full width of the resonances (as
obtained in our resonance model) and the photonic branching ratios, that
we take from the Review of Particle Physics \cite{PDG}. We made
calculations using both the minimal and the maximal values of the
photonic branching ratios given in \cite{PDG}. This provides us with an
estimate of the resulting uncertainty of the dilepton production cross
sections. 

Carrying out the parameter fitting described in Sec.\ \ref{MC}
we arrived at two different
parameter sets describing the experimental data with nearly the same
accuracy. We calculated the dilepton spectrum with both parameter 
sets. From the point of view of dilepton production the main difference
between the two parameter sets is in the baryon resonance Dalitz-decay
channels: using the parameters of our best fit two of the higher
resonance channels, the $N(1520)$ and $N(1535)$ give considerable
contributions to the dilepton spectrum. The contribution of these two
channels exceeds that of the $\Delta(1232)$ at the large dilepton mass
end of the spectrum for large enough beam energies. In the case of the
other parameter set the $N(1520)$ gives a smaller contribution and its
role is partially taken over by the $\Delta(1700)$, which gives a larger
contribution for large dilepton masses. We used the parameters of our
best fit in obtaining all numerical results presented in the rest of the
paper.

\subsection{Comparison with the DLS data}
\label{DLScomp}

\begin{figure}[p]
\begin{center}
\includegraphics[width=7cm]{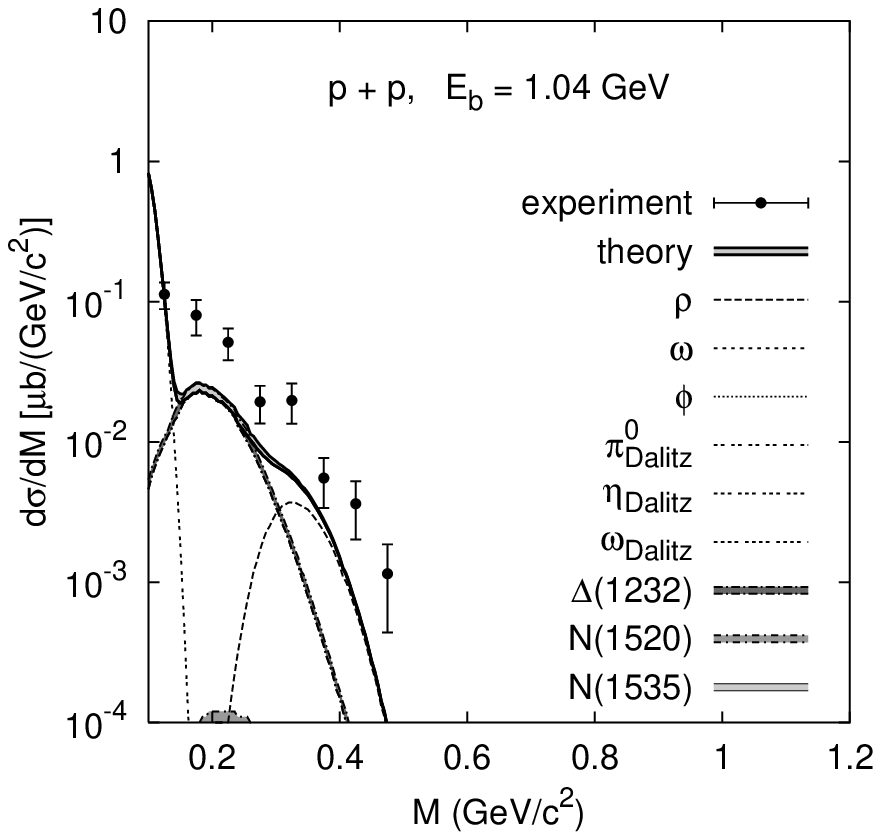}
\includegraphics[width=7cm]{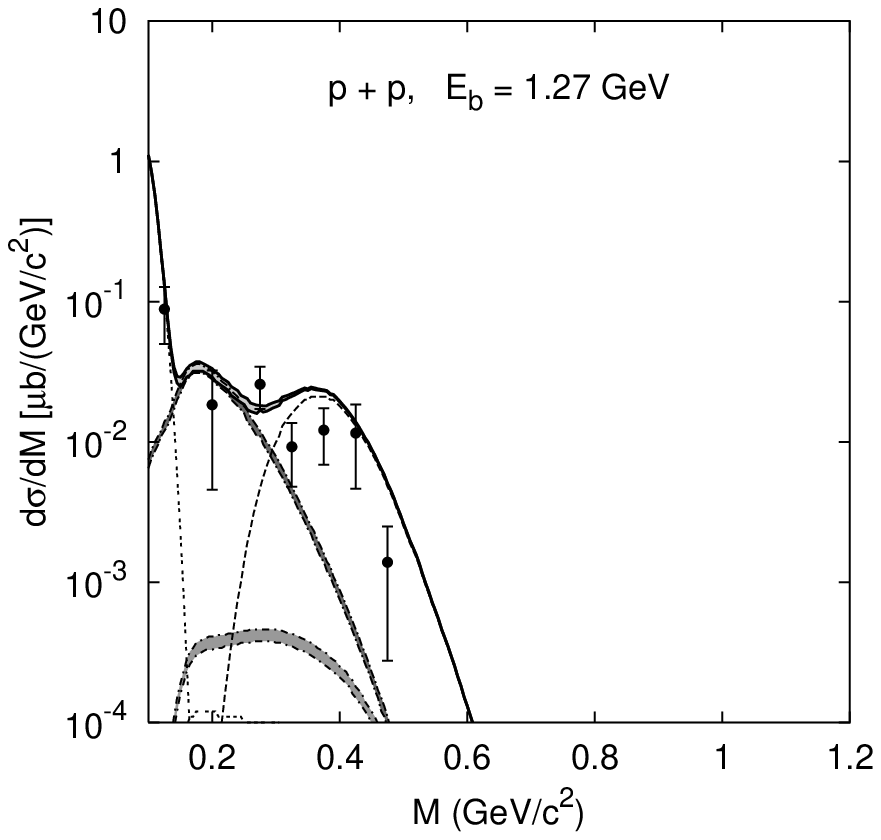}\\
\includegraphics[width=7cm]{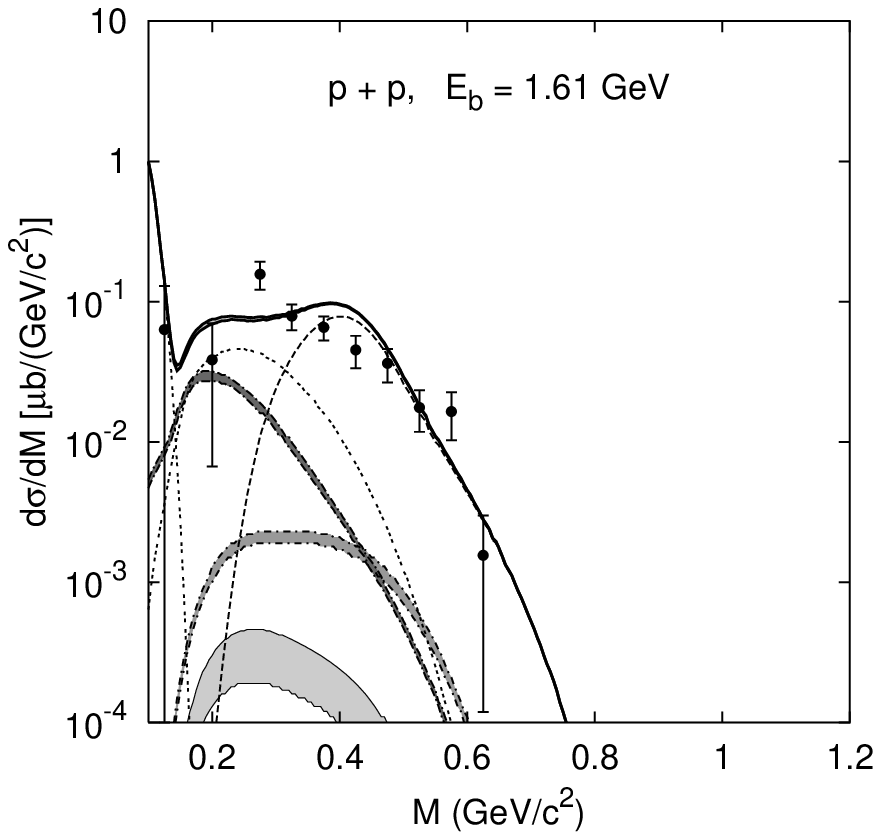}
\includegraphics[width=7cm]{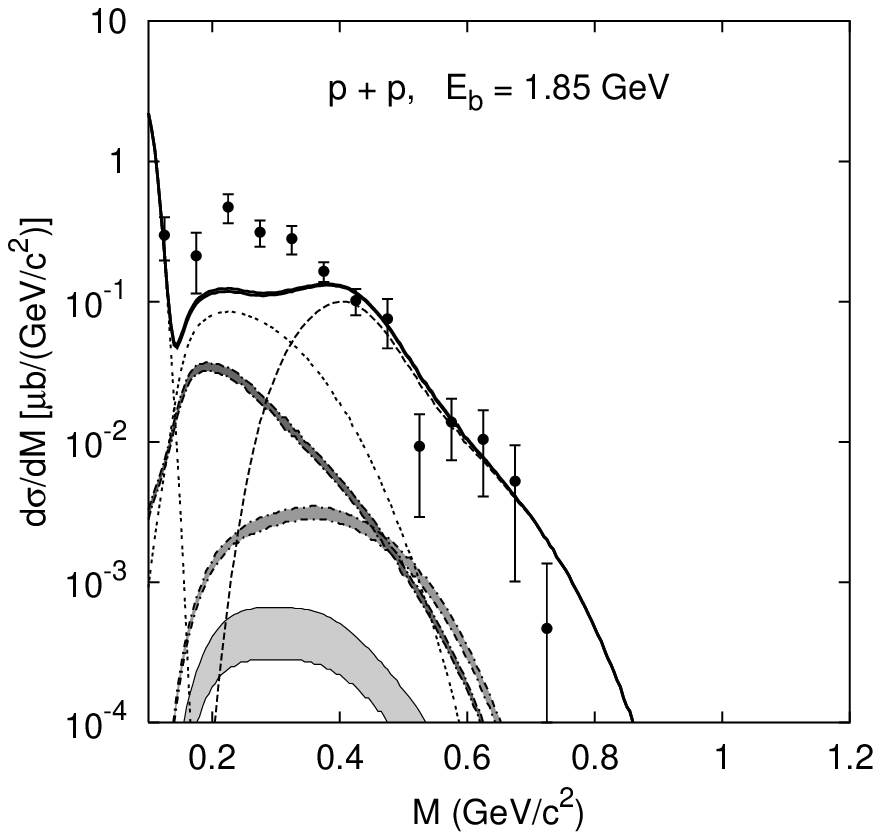}\\
\includegraphics[width=7cm]{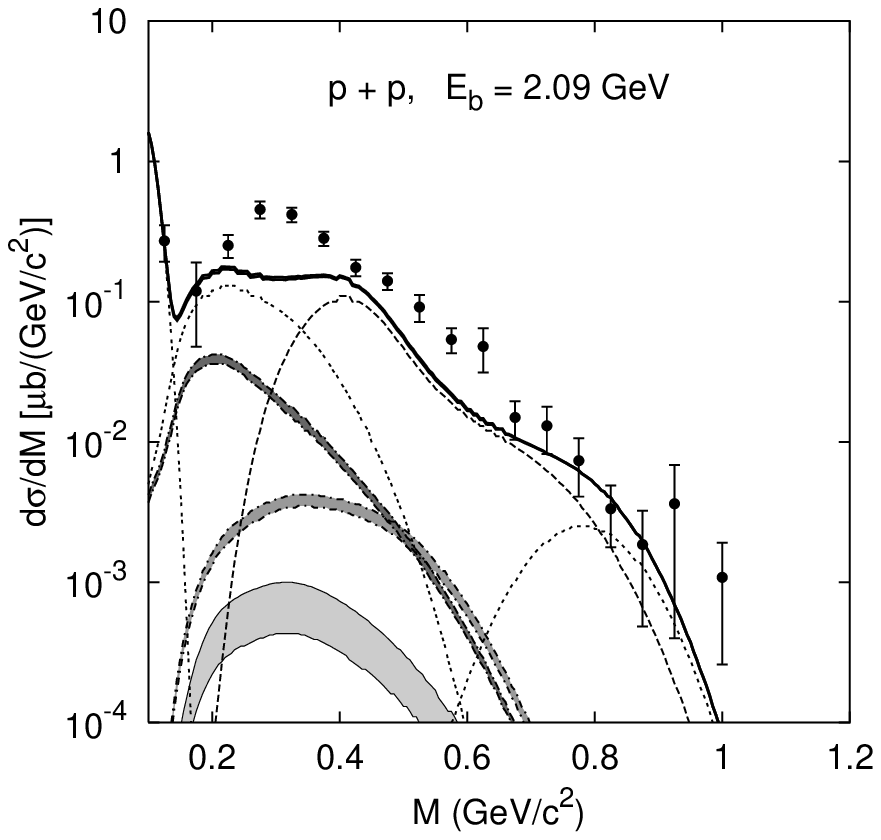}
\includegraphics[width=7cm]{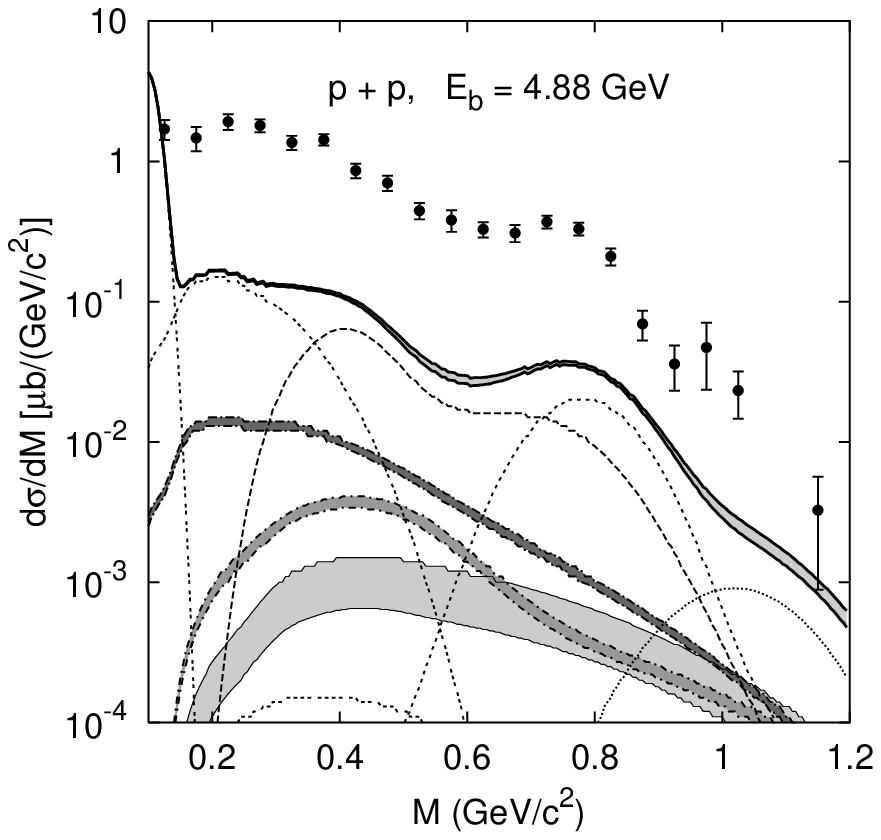}
\end{center}
\caption{\small The dilepton invariant mass spectra for $pp$ collisions
calculated from our model taking into account the DLS filter 
and the finite mass resolution in comparison to the DLS data. 
For the meaning of the various line styles see the explanation 
in the plot for 1.04 GeV beam energy. The solid line represents
the sum of all channels including those that are not indicated in
the plots (i.e.\ Dalitz-decay of the other baryon resonances). In the
case of baryon resonance Dalitz-decays the lower (upper) edge of the
shaded region corresponds to the dilepton production cross section
calculated using the minimal (maximal) value of the photonic branching
ratio given in the Review of Particle Physics. The resulting uncertainty
of the total dilepton spectrum is indicated similarly.}
\label{dlsplots}
\end{figure}

In Fig.\ \ref{dlsplots} we plot the dilepton invariant mass spectra
$d\sigma/dM$ for $pp$ collisions calculated from our model
in comparison to the DLS data \cite{DLS}. We have taken into account the
DLS experimental filter and the finite dilepton mass resolution of the
detector. In the plots contributions
of the following sources are shown: direct dileptonic decay of
$\rho^0$, $\omega$, and $\phi$ vector mesons, Dalitz-decay of 
$\pi^0$, $\eta$, and $\omega$ mesons, and Dalitz-decay of $\Delta(1232)$,
$N(1520)$ and $N(1535)$ baryon resonances. The total cross section is a
sum of all sources including the Dalitz-decay of the other baryon
resonances.

A first glance at Fig.\ \ref{dlsplots} shows that although the resonance
Dalitz-decay channels, esp.\ the $N(1535)$ channel has a large
uncertainty arising from the poor knowledge of the photonic branching
ratios, the resulting relative uncertainty of the full dilepton
production cross section is not dramatic.

The dominant dilepton sources
at the two lowest beam energies are the $\pi^0$ and $\Delta(1232)$ 
Dalitz-decay, and direct dilepton decay of $\rho^0$. At 1.04 GeV
the experimental data are underestimated by the model. Comparison
with results of other models successfully describing the data 
\cite{frankfurt,giessen1,giessen2} suggests
that the reason is probably the underestimation of the $\Delta(1232)$
channel at this energy.

\begin{figure}
\begin{center}
\includegraphics[width=7.5cm]{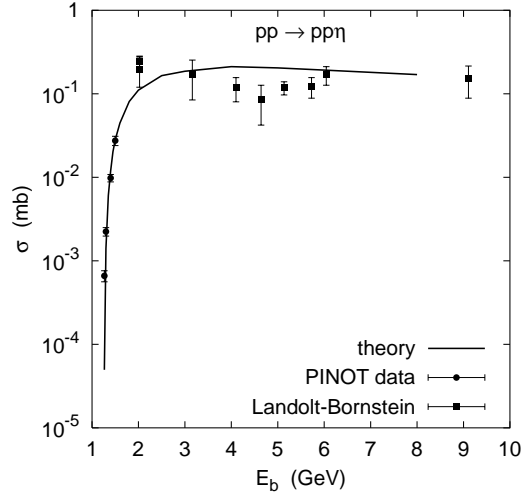}
\end{center}
\caption{\small Cross section of $\eta$ meson production as a function
of the beam energy. The solid line
represents the results from our model while full circles and squares
show the experimental data.}
\label{etacross}
\end{figure}

For beam energies $E_b\ge$ 1.61 GeV the $\eta$ channel gives the dominant
contribution around 0.2 GeV dilepton mass. Figure \ref{etacross}
shows a comparison of the $\eta$ meson production cross section 
deduced from our model to the available experimental data 
\cite{LB,PINOT}. Our model underestimates the two data points at
$E_b =$ 2.02 GeV ($p_{lab} =$ 2.81 GeV). As a result we also underestimate
the dilepton cross section in the $M =$ 0.2 - 0.4 GeV range for the 
beam energies $E_b =$ 1.85 GeV and 2.09 GeV (and, to a lesser extent,
for $E_b =$ 1.61 GeV).

The peak appearing in the dilepton spectrum of the $\rho$ channel at
$M \sim$ 0.4 GeV is a result of two reasons. First, low mass $\rho$
mesons are produced with a high cross section via the decay of
the $N(1520)$ resonance, therefore also the $\rho$ mass spectrum 
has a peak in the same region. Second, close to the two pion threshold
the width of the decay $\rho\to\pi\pi$ becomes small, consequently
the dileptonic branching ratio $BR_{\rho\to e^+e^-}(M)$, that enters
the expression of the dileptonic differential cross section (\ref{Vdilep}), 
rises rapidly as $M$ approaches 2$m_{\pi}$.

In the $M>$ 0.4 GeV region the experimental dilepton spectra are more or less 
well described by the model for the beam energies 
1.27 GeV $\le E_b \le$ 2.09 GeV with the exception of a discrepancy
around $M =$ 0.6 GeV for $E_b =$ 2.09 GeV. The same underestimation
of the data was seen using the model \cite{frankfurt} and a strong
$pp$ bremsstrahlung contribution was named as a possible explanation.

At 4.88 GeV beam energy channels with additional pions in the final
state are known to dominate the dilepton spectrum 
\cite{frankfurt,giessen1,giessen2}. These models use a string fragmentation 
model to describe the DLS data. Our model, since it does not handle
final states with additional pions, looses its validity at this beam
energy. Consequently, our results are an order of magnitude below the
experimental data.

\subsection{Predictions for HADES}
\label{HADESpred}

\begin{figure}[htb]
\vspace*{-0.5cm}
\begin{center}
\includegraphics[width=7.5cm]{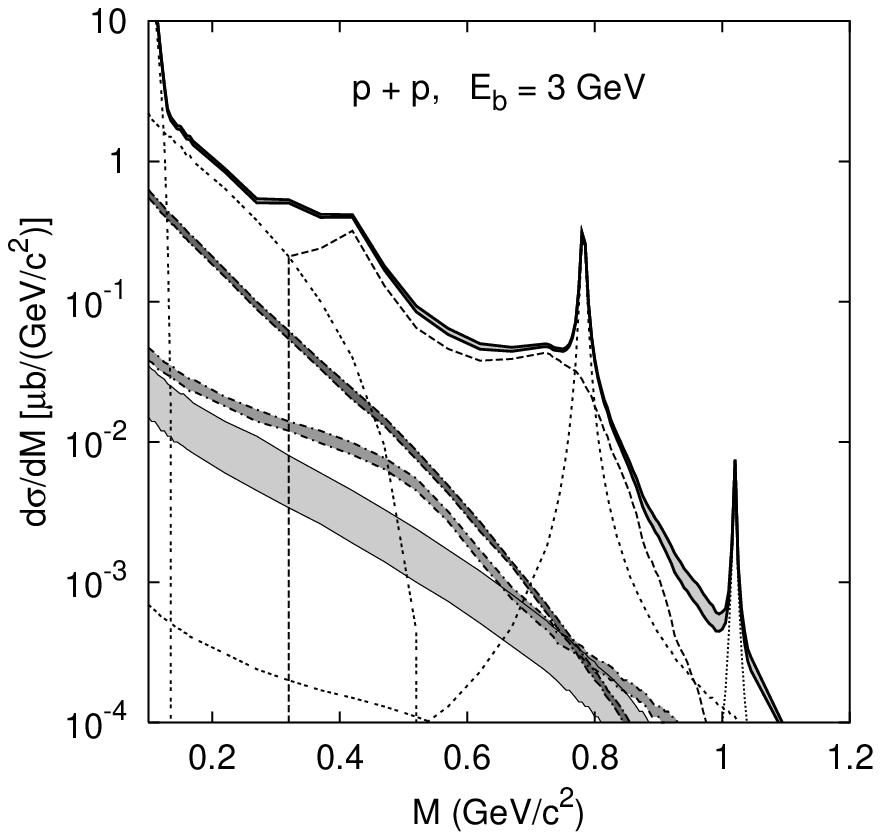}
\includegraphics[width=10cm]{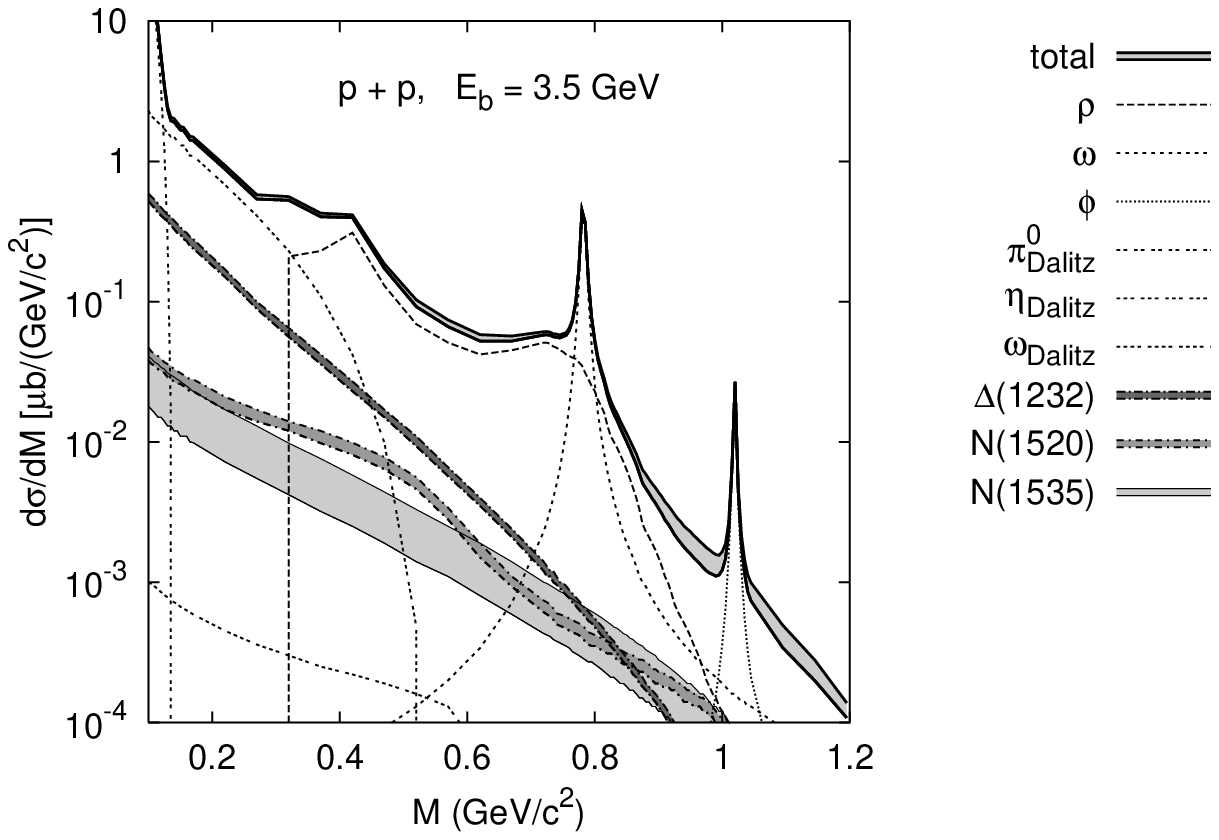}
\end{center}
\caption{
\small Dilepton invariant mass spectrum from proton-proton collisions at
various beam energies. No experimental filter has been used. The
uncertainties of the baryon resonance Dalitz-decay channels and the
total dilepton spectrum are indicated similarly to Fig.\ \ref{dlsplots}.}
\label{dilep}
\end{figure}

Figure \ref{dilep} shows the dilepton invariant mass spectra calculated
from our model for the beam energies $E_b=$ 3 GeV and 3.5 GeV.
Here no experimental filter has been used. In choosing the beam energies
we have taken into account that here we are primarily interested in the
Dalitz-decay channels of higher baryon resonances, which may have
an important contribution in the higher dilepton mass region where
the contribution of the $\Delta(1232)$ is already small, i.e.,
around $M=m_{\phi}$. Therefore, we have chosen beam energies above
the $\phi$ meson production threshold of $E_b=$ 2.59 GeV.
On the other hand, one must not go beyond the beam energy where
the string channel opens and the model looses its validity \cite{frankfurt}.

In Fig.\ \ref{dilep} we see that the dominant dilepton sources
at these energies are the $\pi^0$ and $\eta$ Dalitz-decay and the direct
dilepton decays of vector mesons. Both the $N(1535)$ and the $N(1520)$
channels exceed the $\Delta(1232)$ contribution in the $\phi$ mass
region, but they remain well below the $\phi$ meson
contribution. Consequently, a pronounced peak appears at the $\phi$ mass
in the total dileptonic differential cross section. A similar peak is
seen at the $\omega$ mass. These peaks should be clearly identified by a
high mass resolution dilepton spectrometer such as HADES.
We note that we found the same qualitative features in the case of the
second resonance parameter set, namely, the $\phi$ and $\omega$ peaks
are clearly visible in spite of the fact that the $\Delta(1700)$ channel
exceeds the $\Delta(1232)$ contribution in the $\phi$ mass region.


\section{Summary}

In this paper we investigated the production of dileptons in relativistic
proton-proton collisions using a Monte Carlo model. We introduced as new
channels the Dalitz-decay of a number of higher baryon resonances.
The lack of experimental information about the Dalitz decay of higher
resonances causes a large uncertainty of the calculated dilepton
spectra of these channels. We have given an estimate of the
uncertainties arising from the poor knowledge of the $N\gamma$ branching
ratios of the baryon resonances. Despite the large uncertainties we may
conclude that the Dalitz decay of higher baryon resonances gives a
negligible contribution to the dilepton spectrum.
The dominant dilepton sources are the $\pi^0$ and $\eta$ Dalitz-decay 
and the direct dilepton decays of vector mesons. 

The obtained dilepton spectra
compare well with the experimental data of the DLS collaboration
for beam energies $E_b\le$ 2.09 GeV with some exceptions. 
The most serious disagreement between our results
and the experimental data can be explained by comparing the dilepton
and the $\eta$ production data at the same energies.
We have also given predictions for the HADES detector for beam energies
above the $\phi$ meson production threshold. Our model predicts clear
peaks at the $\omega$ and $\phi$ meson masses, which should be easily
identified by the HADES detector.

Our model can also be used as an
input to a transport model of heavy ion collisions, where in-medium
modifications of vector mesons and baryon resonances can
be studied.


\section*{Acknowledgments}

This work was supported by the National Fund for Scientific Research
of Hungary, grant Nos.\ OTKA T30171, T30855, and T32038, and an MTA-DFG
project. 


\newpage

\begin{thebibliography}{99}
\itemsep=0cm
\bibitem{DLS}
W. K. Wilson et al., Phys. Rev. C {\bf 57} (1998) 1865.
\bibitem{ourpaper}
M. Z\'et\'enyi and Gy. Wolf, accepted for publication in Heavy Ion Phys,
nucl-th/0202047.
\bibitem{LS}
B. E. Lautrop and J. Smith, Phys. Rev. D {\bf 3} (1971) 1122.
\bibitem{PDG}
K. Hagiwara et al., Phys. Rev. D {\bf 66}, 010001 (2002) 1.
\bibitem{BUU}
Gy. Wolf, Heavy Ion Phys. {\bf 5} (1997) 281.
\bibitem{fit}
Gy. Wolf, in preparation.
\bibitem{giessen1}
E. L. Bratkovskaya, W. Cassing, M. Effenberger, and U. Mosel,
Nucl. Phys. {\bf A653} (1999) 301.
\bibitem{Chung}
W. S. Chung, G. Q. Li, and C. M. Ko,
Phys. Lett. {\bf B401} (1997) 1.
\bibitem{N1440}
Gy. Wolf, G. Batko, W. Cassing, U. Mosel, K. Niita, and M. Sch\"afer, 
Nucl. Phys. {\bf A517} (1990) 615.
\bibitem{frankfurt}
C. Ernst, S. A. Bass, M. Belkacem, H. St\"ocker, and W. Greiner,
Phys. Rev. C {\bf 58} (1998) 447.
\bibitem{giessen2}
E. L. Bratkovskaya, W. Cassing, and U. Mosel,
Nucl. Phys {\bf A686} (2001) 568.
\bibitem{LB}
Baldini et al., {\it Landolt-B\"ornstein} vol. 12, (Springer, Berlin, 1987).
\bibitem{PINOT}
E. Chiavassa, G. Dellacasa, N. De Marco, C. De Oliveira Martins, M. Gallio,
P.Guaita, A. Musso, A. Piccotti, E. Scomparin, E. Vercellin,
Phys. Lett. {\bf B322} (1994) 270.
\end{thebibliography}
\end{document}